%% file: bricksmart.tex
\documentclass[sigconf]{acmart}
\AtBeginDocument{%
  }

\setcopyright{acmlicensed}
\copyrightyear{2025}
\acmYear{2025}
\acmDOI{XXXXXXX.XXXXXXX}

\acmConference[CHI'25]{The ACM Conference on Human Factors in Computing Systems}{April 26 - May 1, 2025}{Yokohama, Japan}
\acmISBN{978-1-4503-XXXX-X/18/06}

\acmSubmissionID{4913}

\usepackage{subfigure}
\usepackage{multirow}
\usepackage[utf8]{inputenc} 
\usepackage{graphicx}
\usepackage{makecell}

\input{variables}

\begin{document}

\title{BrickSmart: Leveraging Generative AI to Support Children's Spatial Language Learning in Family Block Play}


\author{Yujia Liu}
\authornote{Both authors contributed equally to this research.}
\affiliation{%
  \institution{Tsinghua University}
  \country{Beijing, China}
}

\author{Siyu Zha}
\authornotemark[1]
\affiliation{%
  \institution{Tsinghua University}
  \country{Beijing, China}
}

\author{Yuewen Zhang}
\affiliation{%
  \institution{Tsinghua University}
  \country{Beijing, China}
}

\author{Yanjin Wang}
\affiliation{%
  \institution{University of Toronto}
  \country{Toronto, Ontario, Canada}
}

\author{Yangming Zhang}
\affiliation{%
  \institution{Wuhan University}
  \country{Wuhan, Hubei, China}
}

\author{Qi Xin}
\affiliation{%
  \institution{Tsinghua University}
  \country{Beijing, China}
}

\author{Lunyiu Nie}
\affiliation{%
  \institution{The University of Texas at Austin}
  \country{Austin, Texas, United States}
}

\author{Chao Zhang}
\affiliation{%
  \institution{Cornell University}
  \country{New York, United States}
}

\author{Yingqing Xu}
\affiliation{%
  \institution{Tsinghua University}
  \country{Beijing, China}
}

\renewcommand{\shortauthors}{Liu and Zha et al.}
\renewcommand{\shorttitle}{BrickSmart}


\input{data/chap00-abstract}


\begin{teaserfigure}
  \includegraphics[width=\textwidth]{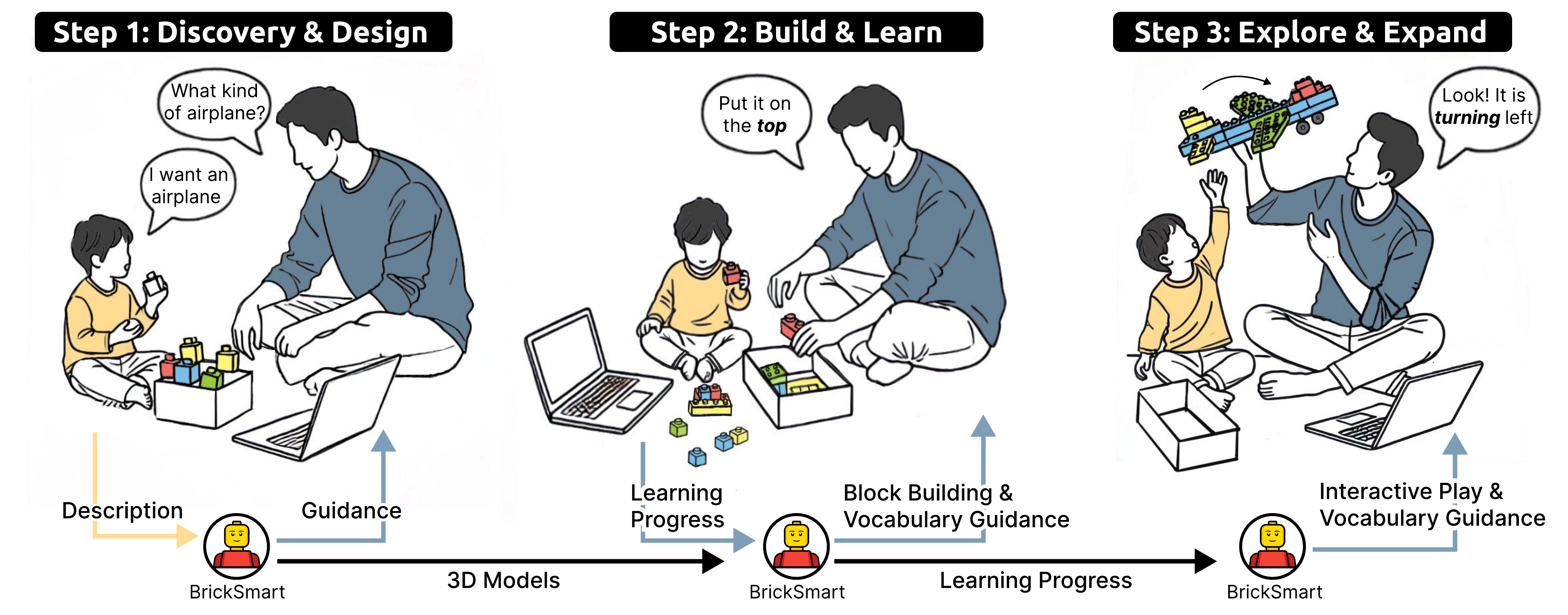}
  \caption{BrickSmart's three-step process: Discovery \& Design, Build \& Learn, and Explore \& Expand. Each step is designed to facilitate parent-child interactions that enhance the child's spatial language and reasoning skills through guided block design, building, and play. Key features include personalized building instructions, active learning progress tracking, and adaptive spatial vocabulary guidance. This process demonstrates a structured method for enhancing early cognitive development in children by integrating generative educational technology into traditional block play.}
  \Description{Enjoying the baseball game from the third-base
  seats. Ichiro Suzuki is preparing to bat.}
  \label{fig:teaser}
\end{teaserfigure}


\maketitle

\input{data/chap01-intro}
\input{data/chap02-related}
\input{data/chap03-guideline}

\input{data/chap04-system_design}
\input{data/chap05-user}

\input{data/chap06-discussion}

\input{data/chap07-conclusion}

\begin{acks}
We would like to extend our profound gratitude to the Future Lab at Tsinghua University for the invaluable support. Additionally, this work was supported by the National Natural Science Foundation of China (Grant No. 62441219) and the Beijing Municipal Science and Technology Project (Grant No. Z231100010323005).
\end{acks}

\bibliographystyle{ACM-Reference-Format}
\bibliography{reference}

\input{data/Appendix}


\end{document}

%% file: variables.tex
\usepackage{balance}       
\usepackage{graphics}      
\usepackage{color}
\usepackage{booktabs}
\usepackage{textcomp}
\usepackage{subcaption}
\usepackage{enumerate}
\usepackage{makecell}
\usepackage{multicol}
\usepackage{multirow}
\usepackage{array}
\usepackage{fdsymbol}
\usepackage{enumitem}
\usepackage{amsmath}
\usepackage{arydshln}
\usepackage{stfloats}
\usepackage{graphicx}
\usepackage{amsthm}
\usepackage{listings}
\usepackage{caption} 
\usepackage{xspace}
\usepackage{csquotes}
\usepackage[bottom]{footmisc}
\usepackage{textcomp}
\usepackage{algorithm}
\usepackage{algorithmic}

\definecolor{DarkGreen}{HTML}{5DAC81}
\definecolor{myred}{RGB}{255,0,0}

\setcitestyle{nocompress}

%% file: data/chap00-abstract.tex
\begin{abstract}



Block-building activities are crucial for developing children's spatial reasoning and mathematical skills, yet parents often lack the expertise to guide these activities effectively. BrickSmart, a pioneering system, addresses this gap by providing spatial language guidance through a structured three-step process: Discovery \& Design, Build \& Learn, and Explore \& Expand. This system uniquely supports parents in 1) generating personalized block-building instructions, 2) guiding parents to teach spatial language during building and interactive play, and 3) tracking children's learning progress, altogether enhancing children's engagement and cognitive development. In a comparative study involving 12 parent-child pairs children aged 6-8 years) for both experimental and control groups, BrickSmart demonstrated improvements in supportiveness, efficiency, and innovation, with a significant increase in children's use of spatial vocabularies during block play, thereby offering an effective framework for fostering spatial language skills in children.

\end{abstract}


%% file: data/chap01-intro.tex
\section{Introduction}

Spatial language, integral to daily life, describes the characteristics and relational dynamics of objects within space, such as ``big/small'' and ``up/down''~\cite{cannon2007system, wu2021early}. 
The development of spatial language during childhood is pivotal as it lays the foundation for advanced spatial cognition~\cite{wu2021early}, logical reasoning~\cite{ferrara2011block}, and mathematical ability~\cite{georges2023importance}. Notably, ages 6-8 represent a critical period for the development of spatial language, during which children make significant gains in understanding and using spatial concepts.~\cite{mishra2003ecology}
Among educational tools, block play, a prevalent family activity, is particularly noted for naturally fostering spatial language and enhancing spatial skills~\cite{ferrara2011block}.

The Vygotskian theory posits that children learn spatial language through block play more effectively with instructional guidance from an experienced partner; without it, play tends to focus purely on entertainment, foregoing educational benefits~\cite{skene2022can}. Research confirms that guided play elicits significantly more spatial language use compared to unstructured or assembly play~\cite{ferrara2011block}. In family settings, parents often serve as facilitators during guided block play, ideally leveraging their familiarity with the child's cognitive level and interests~\cite{ferrara2011block, gregory2003effect, cohen2017block,zha2025colpscaffoldingchildrensonline}. Effective facilitation requires parents to understand spatial language, recognize teachable moments, and provide structured guidance. However, many parents, particularly those from lower socioeconomic backgrounds, lack the necessary skills and knowledge to effectively foster their children's spatial language development through block play.

Human-Computer Interaction (HCI) researchers have explored numerous approaches to support parents in guiding their children's play for educational outcomes like story comprehension~\cite{zhang2023social,zhang2022storybuddy}, language acquisition~\cite{bhatti2021conversational}, computational thinking~\cite{dietz2021storycoder}, and science learning~\cite{xu2020using,zhangBioSketchbookAIassisted2021,zhangObserveItDraw2023}. Notably, Xu et al.~\cite{xu2023mathkingdom} developed a voice agent to enhance the communication between children and parents during video game-based learning of mathematical language. However, parents often find it challenging to guide block play-based spatial language learning, which demands specific skill development. To address these challenges, our study employs Generative AI (GenAI) to provide adaptive, real-time guidance, thereby enhancing parent-child interactions and promoting spatial language development during guided play.

Recent advances in GenAI have facilitated the creation of adaptive, context-aware learning experiences that dynamically cater to a child's specific needs~\cite{du2017learning, labutov2015deep, shakeri2020end}. GenAI's capability to deliver personalized prompts, generate real-time feedback, and provide detailed step-by-step instructions makes it highly suitable for guided block play, where children learn through interactive, hands-on activities. Therefore, this study aims to explore the potential of GenAI in enhancing parent-guided block play, aiming to support the development of children's spatial language skills.


To address the challenges of guiding children in block play-based spatial language learning, we developed BrickSmart, a Generative AI-based system designed to assist parents. BrickSmart provides four core functionalities: Systematize Scaffolding, Personalized Building Instruction Generation, Learning Progress Tracking, and Guide Suggestion Generation. These features work together to create an engaging, adaptive, and interactive learning environment that enhances both educational outcomes and parent-child interactions during block play. 

To evaluate the effectiveness of BrickSmart, we conducted a comparative experiment involving a total of 24 parent-child pairs. The experimental group engaged in guided block play sessions using the BrickSmart system, while the control group did not. Results indicated significant enhancements in spatial language comprehension and usage among children who used BrickSmart, compared to their counterparts in the control group. Moreover, parents in the experimental group reported increased confidence and capability in guiding their children, attributing this to the system's adaptive guidance and real-time feedback. These findings highlight the transformative potential of Generative AI in enhancing guided play as an educational experience.


In summary, the contributions of this study are as follows: 

\begin{itemize} 
 \item Development of BrickSmart, aa system powered by Generative AI that improves parent-child interactions and supports spatial language learning through guided block play;
 \item Conducting a comparative and exploratory user study to assess the effectiveness of BrickSmart in enhancing children's spatial language skills and overall engagement;
 \item Provision of design insights for developing AI systems to assist parents in educational settings, offering guidelines for the future integration of AI in learning environments. 
\end{itemize}

%% file: data/chap02-related.tex
\section{Related work}

\subsection{Spatial Language  Development through Guided Block Play}

Spatial language is essential for describing spatial relationships and characteristics of objects within a specific space, using terms like ``below" and ``behind" to intuitively segment and navigate environments \cite{wu2021early, ferrara2011block}. Early mastery of spatial language is crucial for children as it lays the foundation for advanced spatial cognition, logical reasoning, and mathematical abilities \cite{wu2021early, ferrara2011block, georges2023importance}. For young learners, understanding spatial concepts through language is challenging but fundamental for cognitive development \cite{gentner2013spatial, ferrara2011block}. Research categorizes spatial language into eight types: dimensions, shapes, positions, orientations, quantity, deictics, attributes, and patterns \cite{cannon2007system}, which provide a structured framework for developing spatial language skills.

Block play is an effective and enjoyable method for fostering spatial language development in children \cite{o2004literature}
\cite{yannier2020active}. It provides a hands-on environment that encourages the natural use of spatial vocabulary and cognitive skills \cite{casey2008development, jirout2015building, verdine2014deconstructing}. Studies have shown that both children and parents use more complex spatial language during guided block play compared to free play \cite{ferrara2011block, ramani2014preschool}. In particular, guided play—an approach positioned between free play and direct instruction—has proven effective in promoting spatial language skills. It involves a facilitator, often a parent, setting clear learning objectives while allowing the child to explore actively \cite{weisberg2013guided, kangas2010creative}. By using targeted questions, guiding statements, and heuristic prompts, facilitators can expand children's spatial vocabulary and help them build more complex structures \cite{gregory2003effect, cohen2017block}.

The role of adults in guided play extends beyond providing instructions; it includes offering personalized support based on the child's existing knowledge and interests. Parents, being most familiar with their children's cognitive levels, can deliver more tailored guidance, respond to immediate needs, and provide timely feedback, creating a supportive and motivating learning environment \cite{szechter2004parental, yu2018theoretical,liuWristboundGuanxiJiazu2024a}. Research shows that increased use of spatial vocabulary by adults leads to a corresponding increase in spatial language use by children, highlighting the effectiveness of guided play in developing these skills \cite{skene2022can, pruden2011children}.

In summary, guided block play, supported by effective adult facilitation, provides a balanced approach to enhancing children's spatial language and cognitive development. However, many parents may lack the expertise to provide optimal guidance, presenting challenges in effectively scaffolding learning and maintaining meaningful interactions that promote spatial language skills.

\vspace{-10pt}

\subsection{Generative AI for Children's Learning}
Generative AI is transforming the landscape of children's education by providing personalized, adaptive, and interactive learning experiences. Unlike traditional educational methods, AI-powered tools can dynamically generate content tailored to each child's learning pace, needs, and interests, which is particularly effective for developing both language and cognitive skills \cite{kasneci2023chatgpt, hadi2023survey,zhangMathemythsLeveragingLarge2024,zhangStoryDrawerChildAI2022,zhangStoryDrawerCoCreativeAgent2021}. This ability to adapt in real time is crucial in creating meaningful and engaging learning environments.

AI-driven educational tools like StoryBuddy, which generates personalized narratives based on children's input to promote language development \cite{zhang2022storybuddy}, and Open Sesame, which adapts interactive learning modules to enhance vocabulary and comprehension through contextual learning \cite{lee2024open}, exemplify the potential of Generative AI. These tools help children expand their vocabulary and language comprehension by immersing them in scenarios that require active participation, expression, and contextual understanding~\cite{caiSeeHearTouch2024}. Such adaptability keeps children engaged, promotes deeper learning, and supports their cognitive development by continuously challenging them at the right level. Additionally, O'Malley and Fraser conducted a comprehensive review of learning with tangible technologies, emphasizing how physical interaction with technology enhances cognitive development through hands-on experiences \cite{o2004literature}. In a related study, Yannier, Hudson, and Koedinger explored a mixed-reality AI system for STEM education, demonstrating that learning outcomes improve when active learning extends beyond physical manipulation to include AI-enhanced adaptive feedback \cite{yannier2020active}. These findings highlight how tangible and AI-driven educational tools complement each other, fostering interactive and personalized learning environments.

Moreover, the application of Generative AI in educational settings aligns well with guided play, a method positioned between free play and direct instruction that has proven effective in promoting learning outcomes \cite{weisberg2013guided, kangas2010creative,gong2021holoboard}. In guided play, children benefit from having a facilitator, often a parent or educator, who provides strategic prompts and feedback to guide their learning while allowing them to explore independently \cite{ferrara2011block, ramani2014preschool,zha2024designing}. Generative AI enhances this process by offering adaptive guidance that evolves in response to the child's actions, effectively complementing the role of the facilitator.

Research shows that AI tools can empower parents and educators by providing real-time, adaptive suggestions that enhance their ability to support children's learning even without specialized expertise \cite{zhang2023social,li2023survey,zha2023interdisciplinary,zha2024mentigo}. By generating context-aware prompts and feedback, GenAI helps maintain a balance between child autonomy and necessary support, which is key to effective guided play. Therefore, our study explores the potential of GenAI to enhance guided play by developing an AI-based system that provides adaptive, context-sensitive guidance, empowering parents and educators to support children's spatial language learning better.

\vspace{-10pt}

\subsection{AI Tools for Parent-Guided Play}
Traditional block-building activities often require substantial guidance from parents, who may lack the expertise to provide effective and scientifically grounded support \cite{delacruz2021toy}. As a result, parents often struggle to offer the kind of scaffolded learning that optimizes educational outcomes during these activities. Recent research in the HCI community has explored technological solutions, particularly AI, to provide guided play that enhances parent-child interactions \cite{berson2023exploration, sysoev2022child}. For example, Xu et al. developed a voice-guided game that helps children aged 4 to 7 learn mathematics by building a ``math kingdom." In this game, an AI agent not only encourages children to answer questions but also provides timely feedback, enhancing their understanding of mathematical concepts. The study highlights that AI-driven parent-child games can significantly increase children's interest in learning and improve their mastery of mathematical language, which is crucial for early education \cite{xu2023mathkingdom}.

Advances in natural language comprehension and question generation (e.g., \cite{du2017learning, labutov2015deep, shakeri2020end}) have made it feasible to generate automated guidance that supports children's diverse learning needs while also assisting parents in their facilitative role. This enables interactive question-answering between children and AI systems that can act as collaborative partners to parents. For instance, StoryBuddy, an interactive AI tool designed to facilitate educational goals by engaging both AI and parents, thereby addressing the challenge of maintaining strong parent-child relationships in AI-assisted learning environments \cite{zhang2022storybuddy}. Another study developed a social robot companion that guides and motivates children during storybook reading, enhancing their exploratory learning while enabling parents to play an active role in the process \cite{zhang2023social}. Similarly, StoryCoder uses storytelling as a creative activity, allowing children to modify stories in computational thinking games, promoting creative engagement that can be enriched by parental involvement \cite{dietz2021storycoder}. Conversational agents have also been employed to support children's literacy development \cite{xu2021current}, bilingual language acquisition \cite{bhatti2021conversational}, and science learning \cite{xu2020using}, all of which can be enhanced when parents are included in the AI-driven educational process.

AI tools offer personalized learning experiences by customizing tutorials to each child's unique needs, which can help parents provide more tailored and effective guidance. For example, Open Sesame utilizes a Target Vocabulary Extractor to identify children's vocabulary and then generates storybooks to facilitate targeted vocabulary learning through intervention methods \cite{lee2024open}. Such AI systems can help parents provide structured support that aligns with their child's developmental needs. However, despite these advancements, few studies focus specifically on enhancing parental guidance in children's spatial language development. Therefore, this study aims to use a GAI-based agent to assist parents in guiding children during block play, promoting the development of children's spatial language skills. By leveraging AI, the study seeks to enhance the effectiveness of parental guidance in spatial language learning, empowering parents to be more confident and capable facilitators in their child's educational journey.

%% file: data/chap03-guideline.tex
\section{Design Goals}


Based on insights from prior research on spatial language acquisition in early childhood education (refer to section 2) and reflections from our iterative design process, we have identified four key design goals (DGs) essential for designing a parent-guided, child-centered system to support spatial language learning during block-building activities.\\

\textbf{DG1: Systematize Spatial Language Teaching Through the ``What, When, How" Framework.}
To provide comprehensive and structured spatial language instruction, BrickSmart employs the ``What, When, How" framework:

\textbf{What:} Identifies specific dimensions of spatial language essential for cognitive development, such as spatial relations, shapes, and orientations, which are foundational for logical thinking and mathematics ~\cite{cannon2007system}.
\textbf{When: }Determines the optimal moments to introduce specific spatial terms during play, aligning them with key stages of the block-building activity (e.g., preparation, building, exploration) to enhance contextual learning ~\cite{cohen2017block}.
\textbf{How: }Guides parents in using effective instructional strategies such as scaffolding, interactive questioning, and modeling language to help children understand and use spatial terms~\cite{van2002scaffolding,khoza2022understanding}.

\textbf{DG2: Enhance Engagement through Personalized Learning Experiences.}
Research shows personalized learning experiences can significantly improve children's engagement and learning outcomes, particularly in early education~\cite{almousa2023conceptualization,bang2023efficacy}. BrickSmart provides a personalized starting point where children can choose the content and projects they are interested in, allowing the system to generate corresponding block-building tasks or visual representations of the projects. This approach enables children to focus on goals that interest them, stimulating their curiosity and desire to explore. Simultaneously, the system dynamically adjusts the complexity and type of spatial language based on the child's developmental stage, current language proficiency, and learning progress, ensuring that the content remains challenging but not overwhelming. This combination of personalization and engagement helps children better grasp spatial concepts in a positive learning environment~\cite{bang2023efficacy,grant2014personalized}.

\textbf{DG3: Track Learning Progress and Provide Feedback.}
Monitoring progress and providing adaptive feedback are critical to maintaining effective learning. Research indicates that continuous feedback helps reinforce learning and supports long-term retention of spatial vocabulary~\cite{xu2023mathkingdom,bimba2017adaptive,miller2018children}. BrickSmart integrates a real-time progress-tracking system to evaluate the child's understanding and use of spatial language. This adaptive feature helps parents assess their child's progress and dynamically adjusts the learning content to ensure it aligns with the child's evolving needs.

\textbf{DG4: Encourage Active Parent Involvement.}
Parental involvement is a cornerstone of successful early childhood education, especially in language acquisition \cite{szechter2004parental,yu2018theoretical}. BrickSmart is designed to support parents by providing real-time prompts, examples, and suggestions on how to engage effectively with their child during play. This approach empowers parents to take an active role in their child's learning journey, creating a collaborative environment that fosters growth and mutual engagement.

By following these design goals, BrickSmart provides a structured yet flexible approach to enhancing spatial language development through guided play. The system ensures that both children and parents are fully supported, promoting a meaningful and impactful learning experience.

%% file: data/chap04-system_design.tex
\begin{figure*}[!htb]
    \centering
    \includegraphics[width=0.88\linewidth]{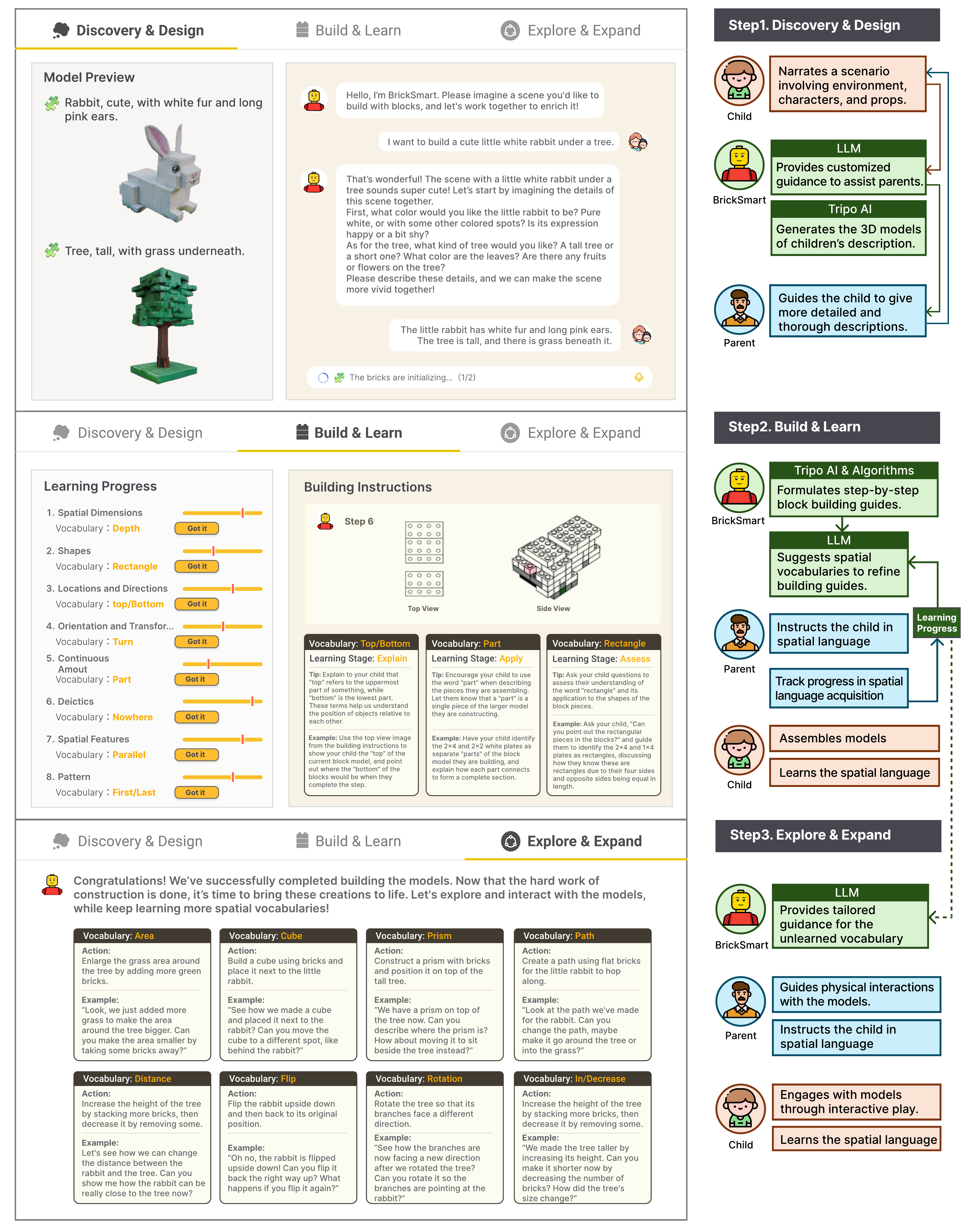}
    \caption{Workflow of BrickSmart. \textbf{Step 1. Discover \& Design:} Children describe their desired scene using voice input, and BrickSmart assists parents to help refine these ideas. The model preview appears on the left. \textbf{Step 2. Build \& Learn:} Parents and children construct the model following the instructions of BrickSmart. Parents are advised to incorporate spatial vocabulary during the building process and track the children's learning progress. \textbf{Step 3. Explore \& Expand:} The assembled models are used for interactive play, where parents introduce more spatial vocabularies to the children as guided by BrickSmart.}
    \label{fig:workflow}
\end{figure*}

\section{System design}
To achieve our design goals, we developed BrickSmart, a GAI-based system designed to help parents guide their children in learning spatial language through block play. BrickSmart operates through three steps—Discovery \& Design, Build \& Learn, and Explore \& Expand (as shown in Fig.~\ref{fig:workflow}). These steps encourage children to engage deeply in building blocks of their interest, learn spatial language systematically, and allow parents to track their children's learning progress. BrickSmart features four core functionalities to support the workflow: Personalized Building Instruction Generation, Systematized Spatial Language Teaching, Learning Progress Tracking, and Guide Suggestion Generation. Each functionality is detailed below.

\subsection{Workflow of BrickSmart}

In order to scaffold children's spatial language learning, we designed systematized steps to guide parent-child collaborative block play. These steps facilitate interaction, promote learning, and adapt to individual learning paces.


\input{table/tab_dimensions}

\textbf{Discovery \& Design (5min):} In this initial step, parents guide their children to describe the scenes they want to build, including elements like environments, characters, and props. BrickSmart processes the child's voice input and generates 1-3 model previews based on the description. The conversation is displayed on the right side of the interface, while the generated previews appear on the left. If the description is too complex, the agent highlights this and prompts adjustments in the dialogue. Once both the parent and child are satisfied with the preview, they proceed to the next step; otherwise, they can regenerate the models. Each generation process takes about one minute, with the entire Discovery \& Design phase limited to five minutes.



\textbf{Build \& Learn (15min):} In this step, parents and children collaboratively construct the chosen scene using the building instructions provided by BrickSmart. During the process, the system guides parents to teach children spatial language terms and concepts. The system provides 74 spatial language vocabulary terms suitable for children aged 6-8, displayed by dimension on the left side of the interface. Each dimension progresses from simpler to more complex terms, with a progress bar showing the child's learning advancement. Parents can evaluate whether the child has mastered a term by asking questions or observing their performance. Once the child demonstrates understanding, the parent clicks ``Got it,'' and the system advances to the next term within the dimension. On the right side of the interface, the current building step and corresponding spatial vocabulary guidance are displayed, dynamically adjusted based on the construction progress and the child's learning status.



\textbf{Explore \& Expand (10 min):} Once the model is built, parents and children interact with it to learn additional spatial language terms in dynamic contexts. During this stage, BrickSmart generates all remaining guiding strategies in one go, based on the completed model elements and the child's progress from the previous step. This step encourages further exploration and reinforces newly learned concepts. Through dynamic interactions with the completed model, children expand their spatial vocabulary while actively engaging with their creations.


Through these three steps, children engage deeply by building something they are interested in, leading to higher engagement. They learn spatial language systematically, and parents can track learning progress and adapt to different learning paces and levels of understanding. The interactive process also serves as a bridge for parent-child bonding through shared building and learning experiences.


\subsection{[DG1] Systematize Scaffolding Spatial Language Learning.}

To scaffold children's spatial language development effectively, we employed the ``What, When, How” framework to design a structured learning process for parent-child guided block play. This approach addresses key questions about what spatial language to teach, when to introduce specific terms, and how to guide parents in facilitating language learning through interactive play.

\textbf{What:} We defined 74 essential spatial language terms relevant to children aged 6-8, organized into key dimensions such as spatial relations (e.g., ``on top," ``next to"), shapes (e.g., ``square," ``rectangle"), and orientations (e.g., ``horizontal," ``vertical"). These terms were carefully selected based on cognitive development theories emphasizing logical reasoning and early mathematical learning~\cite{cannon2007system}. The terms are displayed by dimension on the left side of the system interface, with each dimension linked to a progress bar that reflects the child's learning progress. The vocabulary within each dimension progresses from simpler to more complex terms, supporting incremental learning (as shown in the table.1).

\textbf{When:} The timing of spatial language instruction depends on both the building process and the child's language proficiency. Drawing from \textit{Situated Learning Theory}~\cite{korthagen2010situated}, we designed a three-stage learning sequence to scaffold spatial language learning. In the \textit{Explain} stage, the system introduces new terms with clear definitions and examples relevant to the current building task. In the \textit{Apply} stage, children actively use the terms during construction, guided by system-generated prompts tailored to their progress. Finally, in the \textit{Evaluate} stage, children reflect on their completed models and describe them using the terms they have learned. This staged approach ensures that spatial language is introduced and reinforced at appropriate moments, enabling meaningful and contextualized learning throughout the activity.


\textbf{How:} BrickSmart supports parents by offering step-by-step guidance tailored to the building context and the child's progress. During the ``Discovery \& Design" phase, children describe the scenes they want to build, including environments, characters, and props. The system processes these voice inputs and generates 1-3 model previews displayed on the left side of the interface. If the description is too complex, the agent provides clarification prompts. Parents and children review the previews and either proceed if satisfied or request a new generation, with each generation taking about one minute and the entire phase limited to five minutes. During the ``Build \& Learn" phase, BrickSmart uses Guide Suggestion Generation, powered by LLM), to provide personalized guidance based on the current building step, target spatial vocabulary term, and learning stage (Explain, Apply, or Evaluate). The LLM generates Tips that explain terms and suggest strategies (e.g., ``Introduce rotation by asking the child to adjust the block's orientation'') and Examples offering ready-to-use prompts (e.g., ``Can you try rotating this red rectangular block?''). After assessing the child's understanding, parents can click “Got it” to advance to the next term, ensuring adaptive and personalized learning.


During the building process, parents use Definition Prompts to introduce terms (e.g., ``Let's place the square block `on top' of the long block."), Interactive Prompts to encourage usage (e.g., ``Where should we put this piece to make the tower taller?"), and Reflective Prompts to prompt children to explain their work (e.g., ``Can you tell me which blocks are `next to' the red block?"). Parents can assess whether a child has mastered a term by asking or observing the child's responses. If the child demonstrates understanding, parents click ``Got it” to advance to the next term in that dimension, ensuring adaptive, personalized learning.

\subsection{[DG2]Personalized Building Instruction Generation}

\begin{figure*}[!htb]
    \centering
    \includegraphics[width=\linewidth]{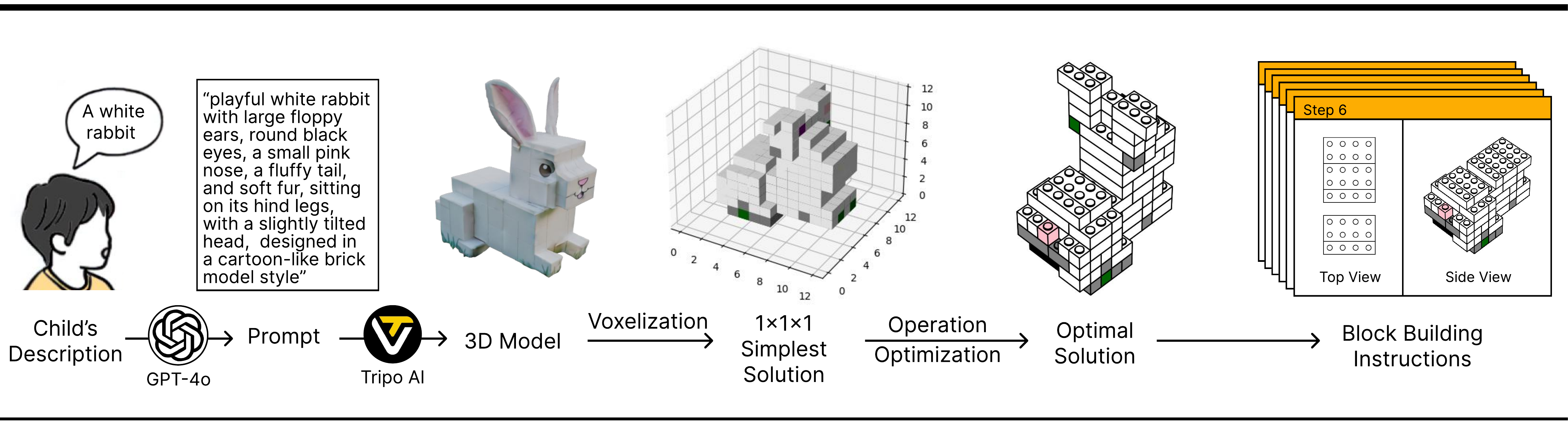}
    \caption{Overview of the personalized building instruction generation pipeline. Including 3D model generation, voxelization, optimization, and generation of formatted instructions.}
    \label{fig:pipeline}
\end{figure*}

BrickSmart transforms a child's description into step-by-step building instructions through an AI-driven pipeline. First, the child's input is refined and used to generate a 3D model using advanced models. Then, the 3D model is voxelized into a matrix of 1x1x1 blocks. An optimization algorithm~\cite{kollsker2021optimisation} refines the block placement to ensure efficiency and simplicity. Finally, the system produces clear, layered instructions with top and side views, making the construction process engaging and accessible.

\textbf{1. 3D Model Generation from Child's Description.}
The pipeline begins with the child's description of their desired structure. This initial text is first revised by GPT-4 to create a more structured and detailed prompt for the 3D model generation. The revised prompt is then fed into the Tripo AI, which uses a generative diffusion model to convert it into a precise 3D object file in \texttt{.obj} format.

\textbf{2. 3D Model Voxelization.}
The 3D model is voxelized into a discrete set of 1x1x1 blocks. This transformation is modeled as a binary programming problem, to cover all required voxel points 
$\mathcal{V}$ using available brick placements $\mathcal{B}$. The relationship between the voxel points and brick placements is defined by the matrix $A_{B,V}$, where each entry represents the coverage of a voxel by a specific brick. The goal is to minimize the total number of bricks used while ensuring complete coverage of the structure:

\begin{equation} \min\ z = \sum_{b\in \mathcal{B}} x_b, \quad  s.t. \sum_{b\in \mathcal{B}} a_{bv}x_b = 1, \quad  \forall v \in \mathcal{V}, x_b \in {0,1}, \forall b \in \mathcal{B} \label{eq1} \end{equation}

\textbf{3. Brick Placement Optimization.}
To improve computational efficiency and interactivity, we use a three-stage heuristic algorithm based on the 'matheuristic' approach \cite{kollsker2021optimisation}.

The first stage decomposes the 3D object into 
1× $X$ strips, processed layer by layer. The second stage applies a 1D-heuristic algorithm to segment co-planar strips into the minimum number of bricks, incorporating gap information G to optimize placement. The cost function is defined as:

\begin{equation}
F_{cost}(b, L) = M(l_b, l_0) + N(L, l_0) + D(b, G) + e
\label{eq4}
\end{equation}

where $b$ is the selected brick, $l_b$ is its length, $l_0$ is the longest available brick, and $L$ is the remaining length of the strip. The multiplicity function $M$ minimizes smaller brick usage, while $N$ optimizes the number of bricks used, and $D(b,G)$ accounts for gap optimization.

In the final stage, adjacent bricks are merged into larger units, reducing the total number of components and simplifying the building instructions.

Detailed information on the optimization process can be found in Appendix \ref{Ap.B}.

\textbf{4. Generation of Formatted Instructions.}
The final step involves generating step-by-step building instructions. These instructions are formatted to include both top and side views, offering clear visual guides for children. The instructions ensure that each construction step is presented in an easy-to-follow, layered format, making the overall building process accessible and engaging.

\subsection{[DG3] Learning Progress Tracking}
We developed the Learning Progress Tracking feature to cater to each child's unique learning pace and provide parents with a comprehensive understanding of their child's progress. This functionality tracks the child's mastery of spatial language across eight key dimensions: Spatial Dimensions, Shapes, Locations and Directions, Orientations and Transformations, Continuous Amount, Deictics, Spatial Features and Properties, and Pattern (as detailed in Table 1).

For each dimension, the system displays a visual progress tracker for specific vocabulary terms, such as ``Big/Little," ``Circle," ``Left/Right," ``Rotate," and ``Increase/Decrease." Parents can easily see which terms have been mastered and which require further practice. This level of detail allows parents to focus on areas that need reinforcement, ensuring a balanced and comprehensive development of their child's spatial language skills.

Moreover, by providing real-time updates on progress, the system helps parents make informed decisions about adjusting the learning plan and tailoring future activities. This dynamic, data-driven approach enables personalized learning paths that adapt to different learning speeds and comprehension levels. It empowers parents to actively engage in their child's educational journey and optimize the learning experience.


\subsection{[DG4] Structured Parental Guidance and Suggestion Generate}

During and after the block-building process, BrickSmart utilizes GPT-4 to generate structured guidance and suggestions that help parents facilitate their child's learning experience. These AI-generated prompts are designed to provide clear, context-sensitive instructions that support both spatial language development and the building activity itself. The prompts guide parents in introducing relevant vocabulary, asking questions to encourage spatial reasoning, and offering constructive feedback to maintain engagement and learning momentum. An example of these prompts is provided in Appendix \ref{Ap.C}. This approach ensures that parents are equipped with the right tools to create a rich, interactive, and educational block play experience.

%% file: table/tab_dimensions.tex
\begin{table*}[h]
\label{tab:dimensions}
\caption{Eight dimensions of spatial language and the corresponding vocabularies \cite{cannon2007system} used in the BrickSmart system.}
\small 
\begin{tabular}{>{\centering\arraybackslash}m{0.5cm}m{11cm}}
\toprule
 & \multicolumn{1}{c}{\textbf{Spatial Language Vocabularies}}\\ 
\midrule
\multicolumn{1}{l}{\textbf{1. Spatial Dimensions}} & 
Big/Little, Long/Short, High/Low, Wide/Narrow, Thick/Thin, Skinny/Fat, Deep/Shallow, Full/Empty, Length, Height, Width, Depth, Volume, Capacity, Area.  \\
\midrule
\multicolumn{1}{l}{\textbf{2. Shapes}} & 
Circle, Square, Rectangle, Triangle, Oval, Semicircle, Polygon, Cube, Sphere, Cylinder, Cone. \\
\midrule
\multicolumn{1}{l}{\textbf{3. Locations and Directions}} & 
At, To/From, On/Off, Ahead/Behind, Left/Right, In/Out, Between/Aside, Opposite, Position, Direction, Distance, Path. \\
\midrule
\multicolumn{1}{l}{\makecell[l]{\textbf{4. Orientations and }\\\textbf{\quad Transformations}}} & 
Forward/Backward, Turn left/right, Upward/Downward, Flip, Rotate, Slide, Clockwise/Counterclockwise. \\
\midrule
\multicolumn{1}{l}{\textbf{5. Continuous Amount}} & 
Whole/Part, All/Half/Fraction, A lot/A little, More/Less, Same/Equal. \\
\midrule
\multicolumn{1}{l}{\textbf{6. Deictics}} & 
Here, There, Where, This, That, Which.  \\
\midrule
\multicolumn{1}{l}{\makecell[l]{\textbf{7. Spatial Features and} \\ \textbf{\quad Properties}}} & 
Line, Curve, Edge, Flat, Bent/d, Point, Acute angle, Obtuse angle, Right angle, Vertical, Parallel.\\
\midrule
\multicolumn{1}{l}{\textbf{8. Pattern}} & 
Increase/Decrease, Before/After, Next/Last, First/Last, Order, Repeat, Pattern.  \\
\bottomrule
\end{tabular}
\end{table*}

%% file: data/chap05-user.tex
\section{User Study}
\begin{figure*}[H]
    \centering
    \includegraphics[width=0.9\linewidth]{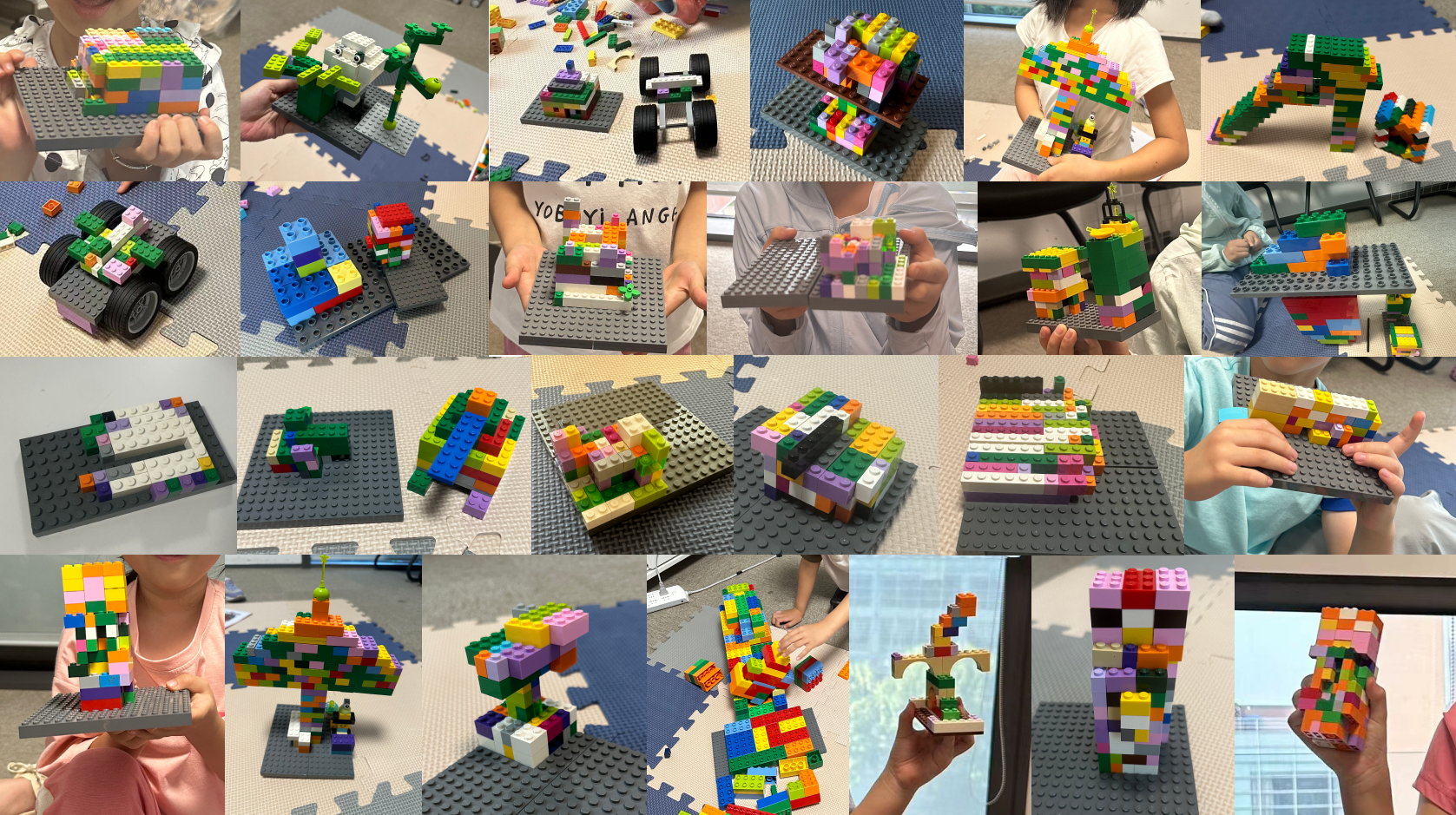}
    \caption{Children's Block Designs. A collection of diverse and creative block constructions by children using the BrickSmart system.}
    \label{fig:userdesigns}
\end{figure*}
We conducted a comparative study to evaluate the effectiveness and usability of BrickSmart in supporting spatial language development during block play with children aged 6 to 8 years. This study aimed to understand how BrickSmart's guided play approach influences children's spatial vocabulary acquisition, engagement, and learning experience. We hypothesize that the personalized spatial language guidance provided by BrickSmart will be perceived as a valuable tool by both children and parents, enhancing children's spatial language skills during block-building activities.

\subsection{Participants}
\input{table/tab_users}
We conducted a user study to evaluate the effectiveness of the BrickSmart system in enhancing spatial language development among children aged 6 to 8 years. A total of 24 parent-child pairs participated in the study, recruited through local community centers and online parent groups. The participants were divided into two groups: an experimental group (12 pairs) using the BrickSmart system and a control group (12 pairs) using traditional block-building methods without system guidance.

To ensure a balanced comparison, children in both groups were matched based on age, gender, and initial spatial language abilities, which were assessed through a pre-screening test. All parents provided informed consent for participation, and ethical approval for the study was obtained from the university's Institutional Review Board (IRB). The study also ensured the safety and comfort of all child participants by having a researcher present during all sessions to monitor their well-being and provide support as needed.

\subsection{Study Procedure and Protocol}
The study employed a between-subjects experimental design, where each parent-child pair participated in one of two conditions: the experimental condition using the BrickSmart system or the control condition using traditional block-building activities. The study procedure was structured into three phases: pre-test (10 minutes), intervention (30 minutes), and post-test (20 minutes), lasting approximately one hour for each pair. The following steps outline the procedure for both groups:

\subsubsection{Experimental Group:}
   \textbf{Pre-Test (10 minutes):} Children completed a spatial language assessment (shown in Appendix \ref{Ap.D}) describing illustrated scenes featuring various spatial relationships and objects. This served as a baseline for evaluating changes in spatial language proficiency. 
    
    \textbf{Intervention (30 minutes):}
\begin{itemize}
    \item \textbf{Step 1:} Children were prompted to articulate scenes, characters, and props they wanted to build using blocks. This initial step encouraged creativity and set the stage for the building activity.
    \item \textbf{Step 2:} Based on the child's input, BrickSmart generated custom building models and provided step-by-step tutorials tailored to the child's ideas. Parents were guided by BrickSmart to use spatial language prompts during the building process, helping the child understand spatial terms (e.g., ``above," ``next to," ``between").
    \item \textbf{Step 3:} After construction, children engaged in interactive play with their models, with BrickSmart offering additional spatial language prompts during movement and play. This helped reinforce vocabulary through context-driven interactions.
    \item \textbf{Step 4:} To conclude, children described their creations in a one-minute narrative, further reinforcing their use of newly learned spatial language terms.
\end{itemize}
    
    \textbf{Post-Test (20 minutes):} The spatial language assessment was repeated using the same illustrated scenes as in the pre-test to measure improvements in spatial language development.

\subsubsection{Control Group:}
    \textbf{Pre-Test (10 minutes):} The control group followed the same pre-test procedure as the experimental group to establish a baseline for spatial language abilities. At the same time, parents attended a 10-minute training session on spatial language definitions, vocabulary, and teaching strategies.
    
    \textbf{Intervention (30 minutes):}
\begin{itemize}
    \item \textbf{Step 1:} Children selected from a set of step-by-step tutorials pre-generated by the BrickSmart system (e.g., rabbit, house, or tree).
    \item \textbf{Step 2:} Children followed the tutorials to construct with their parents' assistance. This process aimed to simulate natural play without the tailored support of BrickSmart, but parents could use their own devices.
    \item \textbf{Step 3:} After completing the construction, children and parents engaged in free play with the models they created.
    \item \textbf{Step 4:} Similar to the experimental group, children described their creations in a one-minute narrative, which aimed to encourage the use of newly learned spatial language terms.
\end{itemize}
    
    \textbf{Post-Test (20 minutes):} A post-test similar to that of the experimental group was conducted to reassess the child's spatial language abilities using the same illustrated scenes.

For both experimental and control groups, in the intervention phase, researchers were present to oversee the process and provide procedural reminders, but did not participate in guiding parent-child interactions. The intervention phase for experimental group was divided into three sub-phases: Discovery \& Design (5 minutes), Build \& Learn (15 minutes), and Explore \& Expand (10 minutes), totaling 30 minutes. For the control group, to maintain consistency in session duration, parents were reminded to use spatial language as frequently as possible during play, rather than moving directly from one building step to the next. This ensured that parent-child conversations about spatial concepts were encouraged, allowing for a comparable evaluation of spatial language development between the two conditions.

\vspace{-5pt}

\subsection{Data Collection and Analysis}

\subsubsection{Pre-test and Post-test}
Children's spatial language proficiency was evaluated using a pre-test and post-test designed with comparable difficulty. Each test included 24 questions: 16 fill-in-the-blank tasks and 8 comprehension tasks, with a total possible score of 48 points. These tests aimed to measure improvements in spatial language use after the intervention. The pre-test and post-test design were developed based on established assessment frameworks in prior research~\cite{gilligan2021aged, farran2016development} and were adapted to align with the cognitive and linguistic characteristics of the target population in this study. Reliability and validity were assessed through pilot testing and expert review, ensuring the tests met standard pre-post evaluation criteria. Internal consistency was measured using Cronbach's $\alpha$, yielding values of 0.81 for the pre-test and 0.78 for the post-test, indicating satisfactory reliability. This robust assessment design enabled a comprehensive evaluation of the intervention's effectiveness in enhancing children's spatial language proficiency.


\subsubsection{Parent Feedback and Usability Measures}
After the study, parents provided feedback on the system by completing three questionnaires: the User Experience Questionnaire (UEQ), which measured user satisfaction across various dimensions; the System Usability Scale (SUS), which assessed overall usability; and the NASA Task Load Index (NASA-TLX)~\cite{hart2006nasa}, which evaluated the perceived workload during the session. Additionally, parents participated in interviews to offer qualitative insights into the system's effectiveness and usability.

\subsubsection{Children's Engagement and Fun}
The Fun Toolkit~\cite{read2002endurability} was used to measure children's engagement and enjoyment, focusing on Endurability, Engagement, and Expectations.

\subsubsection{Video Transcription and Coding}
Video recordings of parent-child interactions during the study were transcribed verbatim to capture the use of spatial language. A coding scheme was developed to identify and categorize spatial language terms and two independent coders analyzed the transcripts. To assess spatial language use, we calculated two key metrics: frequency and density. Frequency was defined as the total number of spatial language terms used by parents and children during the 30-minute interaction session. Density refers to the total number of spatial terms used in 30 minutes, reflecting the intensity of spatial language integration throughout the activity. These measures provided a comprehensive understanding of how effectively spatial language was embedded in dialogues, highlighting the impact of the BrickSmart system on language development during block-building activities.

To ensure the reliability of the coding, we conducted an inter-rater reliability (IRR) test, where two independent researchers coded the dialogues. The overlap between their codes was 89.6\%, indicating high reliability in the analysis process. Additionally, we held joint internal discussions to reach consensus before finalizing the coding and analysis, ensuring consistency and accuracy in data interpretation.



Quantitative data (test scores, UEQ, SUS) were analyzed using t-tests and ANOVA to compare between groups' spatial language improvement and system usability. Qualitative data from interviews and narratives were thematically analyzed to identify key insights on user experience and system effectiveness.

\begin{figure*}[t]
    \centering
    \includegraphics[width=\linewidth]{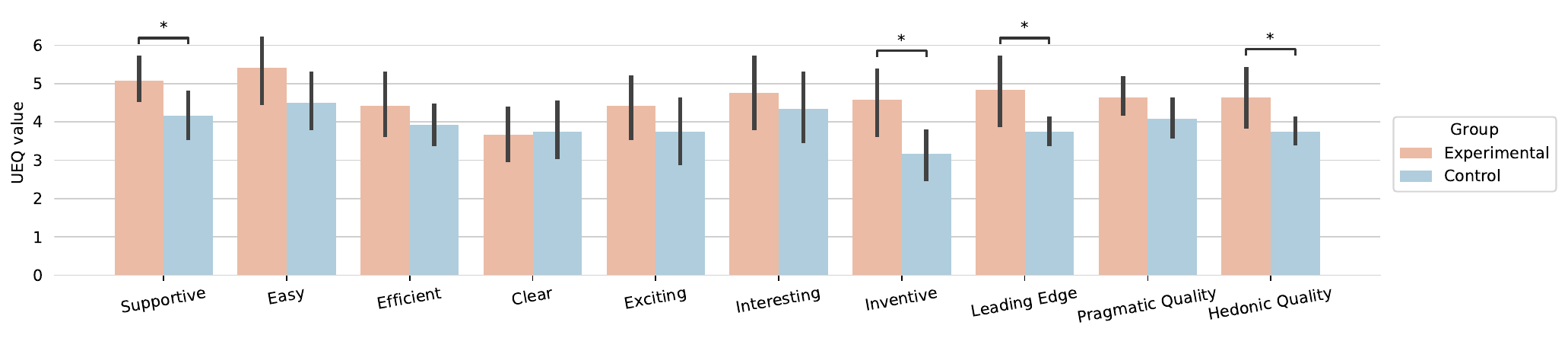}
    \caption{Comparison of UEQ metrics between experimental and control groups. Error bars represent 95\% confidence intervals (CI).  $\ ^*$ stands for $p < 0.05$ and $\ ^{**}$ stands for $p < 0.01$. The same annotation applies to the rest of the paper. }
    \label{fig:ueq}
\end{figure*}

\subsection{Results}
For each statistical analysis, we conducted normality tests on each data set. If the data passed the normality test, we used an independent-sample \textit{t}-test; otherwise, we used the Mann-Whitney \textit{U}-test.

\subsubsection{Evaluation of System Usability}
Statistical analysis confirmed significant differences in several dimensions compared to the control group (as shown in Figure \ref{fig:ueq}). Participants rated BrickSmart significantly better in terms of supportiveness ($t=2.22, p=0.037$), inventiveness ($t=2.55, p=0.018$), leading edge ($t=2.08, p=0.049$), and hedonic quality($t=2.04, p=0.049$)

The System Usability Scale (SUS) results indicate that BrickSmart has high usability, achieving a SUS score of $71.46$. This result suggests that users found BrickSmart intuitive and user-friendly, reinforcing its potential to support spatial language development through guided block play. P8's parent noted that the system allowed them to understand their child's cognitive progress better, while the child developed a preliminary understanding of spatial concepts through play. P3's mother highlighted that integrating vocabulary learning with LEGO building provided valuable educational insights, suggesting that more detailed guidance could further enhance the learning experience. She also mentioned that building with specific learning goals introduced new focus areas, indicating that structured play could lead to more targeted educational outcomes. Overall, these results suggest that BrickSmart not only supports children's learning of spatial language but also provides parents with a framework to better guide and understand their child's learning process, potentially leading to richer educational experiences.

\begin{figure*}[!htp]
    \centering
    \includegraphics[width=0.8\linewidth]{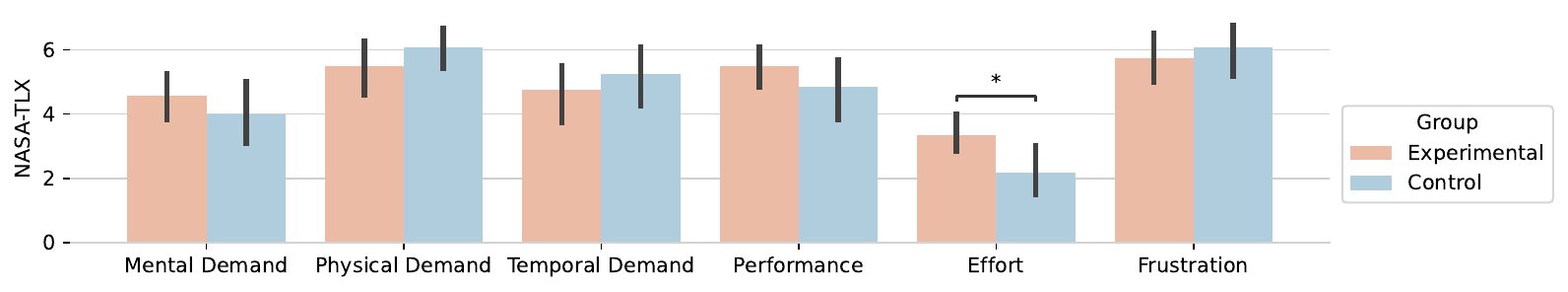}
    \caption{Comparison of NASA-TLX metrics between experimental and control groups. $\ ^*$ stands for $p < 0.05$}
    \label{fig:nasa}
\end{figure*}

In addition, we used the NASA-TLX~\cite{hart2006nasa} to measure parents' perceived workload, as shown in Figure \ref{fig:nasa}. The results indicated that both groups experienced relatively high workload scores, with no significant overall difference between the experimental and baseline conditions. However, on the ``effort" dimension, parents in the experimental group reported significantly higher scores ($t = 2.36, p = 0.027$). This suggests that while BrickSmart effectively supported spatial language learning, its guided interaction process may have required parents to invest additional cognitive and instructional effort.


\begin{figure*}[h]
    \centering
    \subfigure[]{
        \begin{minipage}{3.7cm}
        \centering   
        \includegraphics[width=\linewidth]{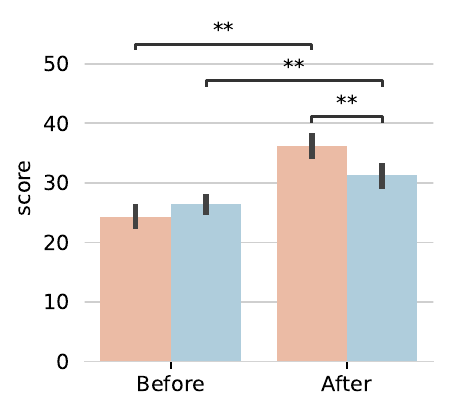}
        \end{minipage}}
    \subfigure[]{
        \begin{minipage}{9.5cm}
        \centering   
        \includegraphics[width=\linewidth]{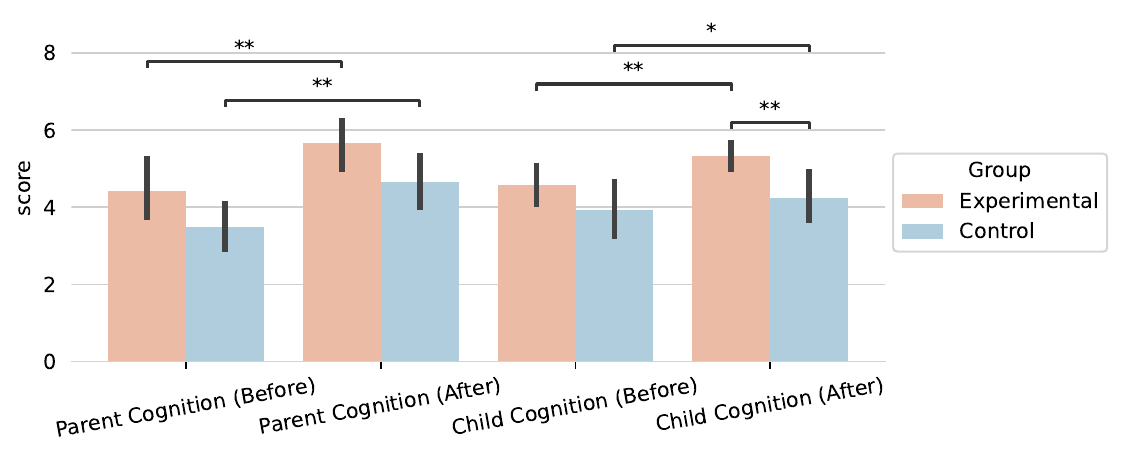} 
        \end{minipage}}
    \caption{Cognitive Score Comparisons: (a) Changes in overall scores before and after the intervention for experimental and control groups. (b) Detailed before and after scores for parent and child cognition in both groups, with significant differences marked.}
    \label{fig:beforeafter}
\end{figure*}

\subsubsection{Children's Spatial Language Skill Improvement and Performance}
We evaluated the impact of the BrickSmart system on children's spatial language skills using three approaches: pre- and post-test assessments of children's knowledge, parental evaluations of their children's progress, and analysis of video transcriptions to measure the density and frequency of spatial language use in parent-child dialogues. These data provide a comprehensive understanding of how the system enhances spatial language development.

\textbf{Student Test Results on Spatial Language Skill}
As shown in Figure \ref{fig:beforeafter}, it presents the Spatial Language Questionnaire results, which measured children's spatial language skills before and after the study for both groups. Statistical analysis shows that both groups significantly improved in spatial language skills from pre-test to post-test (Experimental: $t=10.24, p<0.001$. Control: $U=1.0, p=0.004$). However, the experimental group showed a significantly greater improvement, with an average increase of 49.0\%, compared to the control group’s 18.23\% improvement. The post-test scores of the experimental group were markedly higher ($U=127.0, p=0.002$), indicating that the BrickSmart system was more effective in enhancing children's spatial language abilities than the traditional approach.

\textbf{Parental Assessment of Children's Spatial Language Improvement}
Figure \ref{fig:beforeafter} shows the before-and-after evaluations from parents regarding their children's spatial language development. The results reveal that both groups exhibited an increase in perceived spatial language skills after the study ($all ~p<0.05$), but the increase was more pronounced in the experimental group. Parents in the experimental group reported significantly higher improvements in their children's spatial language cognition compared to the control group ($U=117.5, p=0.007$). This suggests that the BrickSmart system not only enhances children's spatial language skills but also leads to noticeable improvements as perceived by their parents.

The parents' interview feedback further supports these quantitative findings. P8's mother said, \textit{``Spatial language is a crucial part of cognitive training for children, primarily acquired through natural learning and school materials. During this activity, I noticed my child learning about different categories of spatial language and gaining new educational insights."} Similarly, P3's parent noted that the experiment helped their child recognize the importance of spatial language, but also mentioned that some new terms introduced during the study were challenging to encounter in regular block-building activities.

\begin{figure*}[h]
    \centering
    \includegraphics[width=\linewidth]{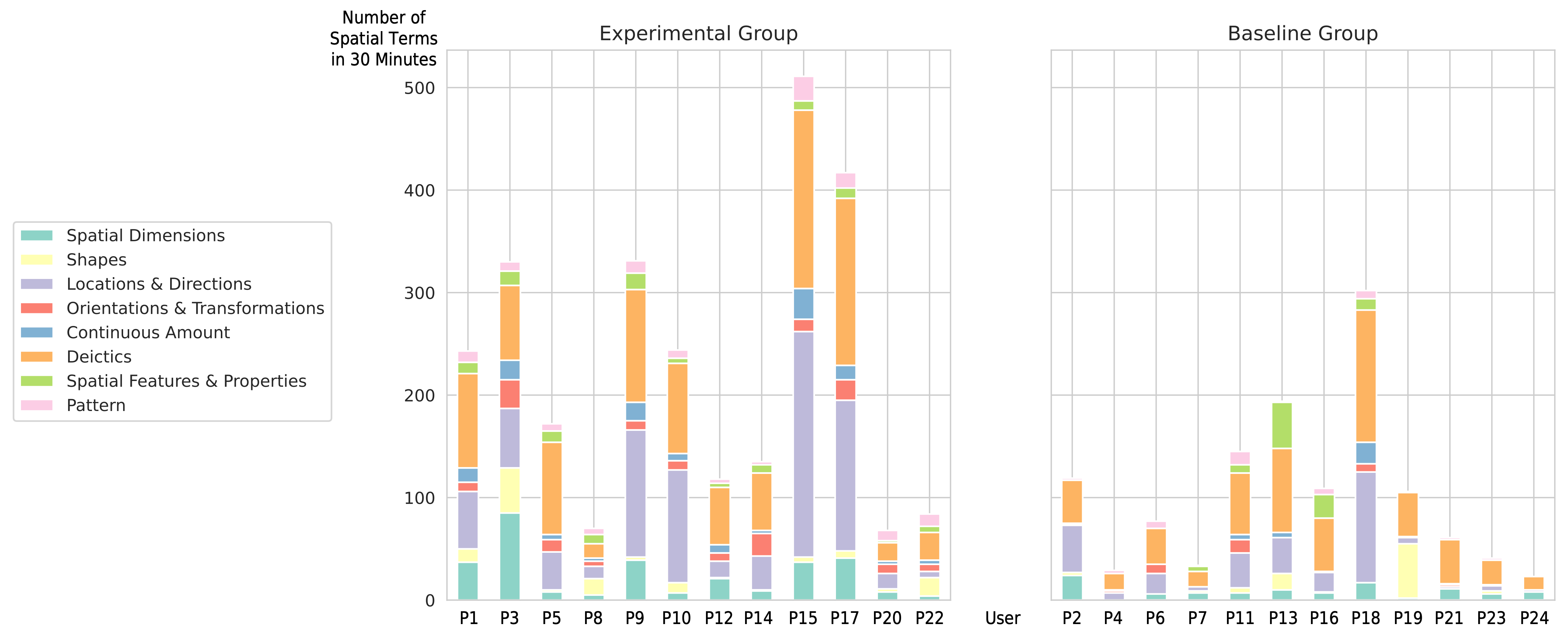}
    \caption{Comparison of category frequencies in spatial language vocabularies' presence. The experimental group with BrickSmart demonstrates higher overall frequency and more comprehensive coverage across dimensions. }
    \label{fig:freq}
\end{figure*}

\textbf{Analysis of Spatial Language Usage from Transcribed Dialogues}
Based on the transcription and coding of videos recorded during the study sessions, we analyzed the density and frequency of spatial language used in dialogues between children and parents. As shown in Figure \ref{fig:freq}, the results show a comparison of the frequency of different categories of spatial language vocabularies used in parent-child dialogues between the experimental group and the baseline group. The experimental group shows a notably higher frequency of spatial language terms across several categories. For example, in the spatial dimensions category, the experimental group used these terms over 200 times, while the baseline group used them less than 100 times. Similarly, the locations and directions category shows a significant increase in the experimental group, with terms used nearly 300 times compared to around 150 times in the baseline group.

The orientations and transformations category also saw a marked increase in the experimental group, reflecting a more comprehensive use of terms describing spatial orientation changes. These were used nearly 150 times compared to less than 50 in the baseline. Additionally, terms related to deictics (e.g., ``here," ``there") and continuous amount (e.g., ``more," ``less,") were more frequently used by the experimental group, highlighting their engagement with more complex spatial concepts. Qualitative feedback from parents supports these findings. P15 noted that their child became more accurate in describing spatial orientations. P22 mentioned, \textit{``I noticed that when positioning objects during spatial configuration, along with hands-on manipulation, my child's spatial abilities seemed to improve."} Similarly, P14 observed an enhancement in their child's spatial language skills, stating, \textit{``I could see a deeper understanding of spatial concepts when my child compared overall images with bird's-eye views."} Moreover, we observed that participants in the experimental group used unique spatial terms that were not commonly found in the baseline condition. In the Location and Direction dimension, P8 and P20 frequently used terms like ``suspended'', ``overlapping'', and ``intersecting''. Similarly, in the Spatial Features and Properties dimension, terms such as ``tilted'' were used by P9, while words like ``notched'' and ``hollow'' appeared frequently, especially from P9, P12, P15, and P17. These instances suggest that the system encouraged children to adopt more precise and descriptive spatial language, reflecting a deeper understanding of complex spatial relationships.

These findings suggest that the BrickSmart system promoted a richer and more varied use of spatial language, providing children with more opportunities to practice and internalize these concepts during block-building activities. This increased diversity and frequency in spatial language usage in the experimental group compared to the baseline group illustrates the system's effectiveness in enhancing spatial language development through guided interaction.

\subsubsection{Children and Parent Engagement Across the System}

\begin{figure}[h]
    \centering
    \includegraphics[width=\linewidth]{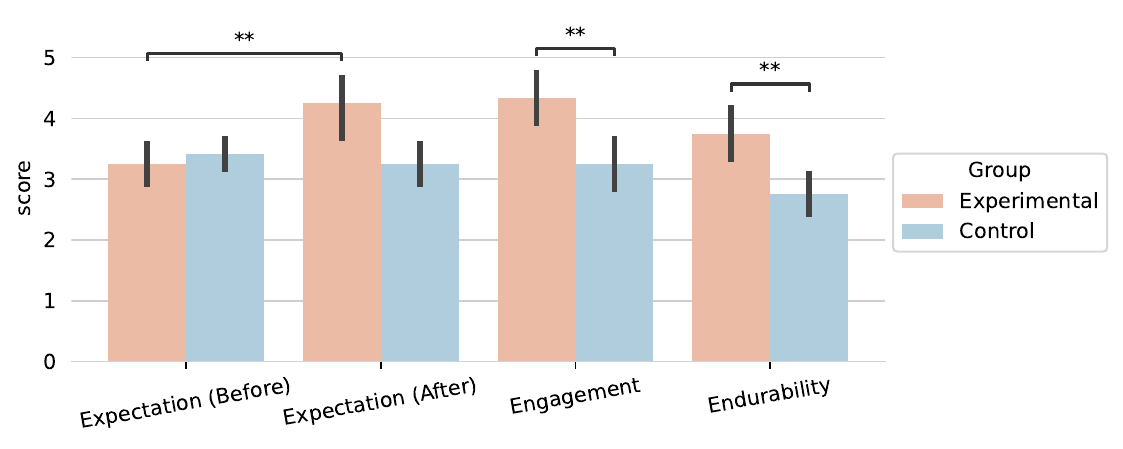}
    \caption{The engagement and fun
assessment results. Notable increases in post-use expectations and engagement were observed in the experimental group, as well as a marked increase in endurability. }
    \label{fig:fun}
\end{figure}

Figure \ref{fig:fun} illustrates the results of the engagement and fun assessments for children across different dimensions: expectation, engagement, and endurability. The results show that while both groups had similar expectations before the study, the experimental group reported significantly higher expectations after using BrickSmart system ($U=28.5, p=0.008$). Additionally, engagement levels were notably higher in the experimental group ($t=3.46, p=0.002$), indicating that the interactive features and guided prompts in BrickSmart helped sustain children's interest and focus throughout the activities. In terms of endurability, or the desire to continue using the system, the experimental group also scored higher than the control group ($t=3.55, p=0.002$). 

Parents in the control group frequently mentioned feelings of achievement (P11, P16, P18, P19, P23), while those in the experimental group emphasized different aspects. For example, P5 noted that BrickSmart aligned well with their child's interests, and P22 mentioned that their child was more engaged because the activities related to recent experiences and their interest in LEGO. P15 highlighted that the system's tutorials were varied and flexible, unlike rigid, rule-based instructions, allowing for greater creativity.

An interesting observation was that while both groups had elements of creative freedom—such as choosing different colors when specific ones were unavailable—the experimental group demonstrated a higher degree of flexibility. Children in the experimental group were more likely to alter shapes or incorporate additional LEGO pieces, suggesting that the initial guided storytelling in Step 1 might have sparked their creative instincts. Another interesting observation is that some parents in the control group, after learning about spatial language, independently adopted more effective and flexible strategies, such as storytelling, making analogies, and others. However, they relied heavily on the parents' higher cognitive abilities, adaptability, and educational approaches.


Parent feedback also highlighted the system's role in enhancing parent-child interaction. While some parents in the control group, like P21, felt that good instructions reduced the need for parental involvement, others, such as P23, observed that children mostly worked independently with minimal parental intervention. In contrast, parents in the experimental group, such as P22, noted frequent interactions with their children, offering immediate encouragement when they encountered difficulties. P8 pointed out a balanced experience, noting that while the system's guidance was helpful, having specific goals also created a sense of urgency.

Most parents from both groups acknowledged that collaboration during the activities helped strengthen their bond with their children. For instance, P7 (from the experimental group) mentioned that having a clear guide and a goal that interested the child made collaboration easier and increased the child's participation. This aligns with our earlier claim that the system can serve as a ``bridge" for parent-child communication, enhancing engagement and cooperation during guided play.

Overall, these findings suggest that BrickSmart not only enhances children's engagement and desire to learn but also fosters meaningful parent-child interactions, making it an effective tool for guided play that supports both educational and relational outcomes.


%% file: table/tab_users.tex
\begin{table*}[h!]
\centering
\small
\begin{tabular}{c|c|c|c|c|c|c|c}
\toprule
\textbf{User} & \textbf{Group} & \textbf{Gender} & \textbf{Parent} & \textbf{Age} & \textbf{Block Play Frequency} & \textbf{Accompany} & \textbf{Number of Models}\\ \hline  
P1 & Experimental & Boy & Father & 8 & Often & With Parent & 2\\ \hline
P2 & Control & Boy & Mother & 6 & Occasionally & Alone & 2\\ \hline
P3 & Experimental & Boy & Mother & 6 & Often & Alone & 1\\ \hline
P4 & Control & Girl & Mother & 7 & Often & Alone & 2\\ \hline
P5 & Experimental & Boy & Mother & 6 & Occasionally & With Parent & 2\\ \hline
P6 & Control & Boy & Mother & 6 & Often & With Parent & 3\\ \hline
P7 & Control & Boy & Mother & 6 & Often & With Parent & 2\\ \hline
P8 & Experimental & Boy & Mother & 6 & Often & Alone & 1\\ \hline
P9 & Experimental & Boy & Father & 6 & Often & With Parent & 2\\ \hline
P10 & Experimental & Boy & Mother & 6 & Often & Alone & 1\\ \hline
P11 & Control & Girl & Grandmother & 6 & Often & With Parent & 2\\ \hline
P12 & Experimental & Girl & Mother & 6 & Occasionally & With Parent & 1\\ \hline
P13 & Control & Boy & Mother & 6 & Often & With Parent & 1\\ \hline
P14 & Experimental & Boy & Mother & 6 & Occasionally & With Parent & 1\\ \hline
P15 & Experimental & Girl & Mother & 6 & Occasionally & With Parent & 1\\ \hline
P16 & Control & Girl & Father & 6 & Occasionally & With Parent & 2\\ \hline
P17 & Experimental & Boy & Mother & 6 & Often & Alone & 1\\ \hline
P18 & Control & Girl & Mother & 6 & Occasionally & With Parent & 1\\ \hline
P19 & Control & Girl & Mother & 7 & Occasionally & With Parent & 2\\ \hline
P20 & Experimental & Boy & Father & 7 & Occasionally & With Parent & 1\\ \hline
P21 & Control & Boy & Mother & 8 & Often & Alone & 3\\ \hline
P22 & Experimental & Boy & Mother & 7 & Occasionally & Alone & 2\\ \hline
P23 & Control & Boy & Mother & 6 & Often & Alone & 2\\ \hline
P24 & Control & Girl & Mother & 6 & Often & Alone & 1\\ \bottomrule

\end{tabular}
\caption{Participant Overview in BrickSmart Study. This table lists the demographics of children in both experimental and control groups, detailing their age, gender, parent involvement, and the frequency of their prior engagement with brick-based activities.}
\label{tab:Users}
\end{table*}

%% file: data/chap06-discussion.tex
\section{Discussion}

This paper explores the potential of BrickSmart, an AI-driven system, in enhancing children's spatial language development through guided block play, contributing to the growing body of research on AI-supported educational tools and human-AI collaboration in learning contexts. In this section, we discuss the insights gained from designing and evaluating BrickSmart, the implications for future AI-based educational tools, the limitations of our current study, and potential avenues for future research.

\subsection{Integrating AI for Spatial Language Development}
The integration of AI-driven guidance in parent-child learning environments raises critical questions about balancing structured support with meaningful human interaction. There are varying perspectives on the role of AI in augmenting parent-child interactions for educational purposes. Some studies suggest that AI can effectively scaffold learning by providing personalized and context-sensitive feedback, thereby enhancing the educational experience \cite{chen2021twenty,bilad2023recent,fernandez2024evaluating}. However, others caution against relying solely on AI systems, emphasizing the importance of human facilitation and the need for tools that support rather than replace parental involvement in children's learning \cite{berson2023exploration, zhang2022storybuddy,sysoev2022child}. Our work aligns with the latter perspective by presenting BrickSmart as a system that integrates AI support while enhancing the role of parents as active facilitators during guided play.

Several design elements likely contributed to BrickSmart's effectiveness in enhancing children's spatial language development. Its structured workflow, consisting of Discovery \& Design, Build \& Learn, and Explore \& Expand phases, helped parents manage the activity while reducing cognitive load for both themselves and their children. Additionally, personalized instruction, which adapted building prompts based on each child's progress, likely increased engagement by providing targeted, context-aware support. Finally, vocabulary-specific prompts, introducing spatial terms through interactive examples and contextual explanations, likely improved children's comprehension and retention of key spatial language concepts. Future research could explore how these features individually and collectively impact learning outcomes. Applying BrickSmart in real-world contexts, such as homes or classrooms, would provide valuable insights into its scalability, adaptability, and potential to foster collaborative, context-aware learning environments.

\subsection{Design Considerations and Implications for AI-Supported Children's Spatial Language}

\subsubsection{\textbf{Adaptive Scaffolding Through Personalized Instructions.}}
The Personalized Building Instruction Generation in BrickSmart serves as adaptive scaffolding, enabling tailored guidance based on each child's unique learning pace and needs. This aligns with prior research suggesting that personalized, responsive learning environments can significantly enhance engagement and learning outcomes \cite{xu2023mathkingdom, zhang2022storybuddy}. BrickSmart dynamically adjusts building instructions to the child's preferences and current level of understanding, making the block-building process more engaging and accessible. This personalized approach ensures that children are neither overwhelmed by complexity nor disengaged due to a lack of challenge. The adaptability of AI in this context demonstrates its potential to provide just-in-time scaffolding that supports children's learning while respecting their autonomy. Future educational tools should continue to explore how adaptive AI can cater to diverse learner profiles, encouraging exploration and creativity within structured learning environments.

\subsubsection{\textbf{Context-Aware Parental Guidance for Enhanced Engagement.}}
BrickSmart also incorporates Guide Suggestion Generation, which provides real-time, context-sensitive prompts for parents to facilitate deeper engagement during play. This feature addresses a critical need in guided learning: empowering parents with the tools to effectively scaffold their child's language development without needing extensive pedagogical expertise \cite{berson2023exploration, sysoev2022child}. By offering suggestions that guide parents on how to introduce spatial terms or ask thought-provoking questions, BrickSmart supports richer parent-child interactions. This aligns with findings emphasizing the importance of active adult participation in enhancing children's learning experiences \cite{zhang2023social}. However, balancing AI suggestions with parental autonomy is crucial to ensure the interactions remain natural and meaningful. Future systems should refine this balance, perhaps through customizable guidance options that allow parents to tailor the AI's input based on their comfort and the child's responsiveness.

\subsubsection{\textbf{Balancing System Autonomy and Human Agency.}}
Our findings highlight the importance of balancing system autonomy with human agency to preserve natural, creative parent-child interactions. While AI-driven prompts structured learning sessions effectively, over-reliance on system guidance risked reducing spontaneous, creative exchanges during play. This observation aligns with prior research emphasizing that excessive system automation can undermine user autonomy, hinder critical thinking, and suppress exploration in learning environments \cite{johnson2012autonomy,gorissen2015autonomy,alzahrani2017exploring}. For example, studies on adaptive learning systems have shown that overly prescriptive feedback reduces engagement, while user-controlled interactions foster greater satisfaction and more dynamic learning experiences \cite{lin2024design,aldila2024effectiveness,conati2021toward}. To address this, future designs should incorporate a parent-driven mode, enabling parents to select when and how the system provides prompts. This approach could include manual activation options or adjustable intervention levels, empowering parents to adapt AI support to their specific needs while maintaining flexible, meaningful interactions. As highlighted in prior work on co-creative and collaborative learning environments \cite{griffith2021discovering,kangas2010creative}, such systems should adapt to varying levels of parental involvement, encouraging spontaneous and context-sensitive interactions while still benefiting from structured, AI-driven guidance.

\subsection{Limitation and Future Work}
While BrickSmart demonstrates potential in enhancing spatial language learning, several limitations should be addressed in future research. First, the study's limited sample size and demographic diversity may restrict the generalizability of findings. Future studies should recruit broader, more diverse participant groups to evaluate the system's effectiveness across various contexts. Additionally, the study focused on short-term learning outcomes, leaving the long-term impact on spatial language development unclear. Longitudinal research could explore whether skills gained through guided play are retained and transferable to other domains, such as academic performance or real-world problem-solving. 

The system's structured task flow, while effective, differs from the more flexible nature of the control condition, which may have acted as a confounding factor. Future research should incorporate structured control conditions to better isolate the system's contributions and consider gain score comparisons for a clearer evaluation of individual learning improvements. Additionally, the reliance on detailed textual instructions may increase parents' cognitive load, especially for those less familiar with spatial language concepts. Enhancing the system with multimodal aids, such as animations, videos, or highlighted keywords, could make instructions more accessible and engaging. Futhermore, for parents with higher cognitive abilities, we should encourage more autonomy, allowing them to use a broader range of flexible, self-generated strategies. Addressing these limitations could further strengthen BrickSmart's usability and effectiveness, making it a more adaptable and scalable tool for supporting children's spatial language development across diverse educational settings.

%% file: data/chap07-conclusion.tex
\section{Conclusion}

This study introduced BrickSmart, a GenAI-driven system designed to support parents in enhancing children's spatial language development through guided block play. By leveraging natural language processing and interactive prompts, BrickSmart facilitates meaningful parent-child interactions, creating a more effective learning environment for spatial language acquisition. Our findings from both controlled and exploratory studies demonstrate that BrickSmart significantly improves children's spatial language skills and engagement compared to traditional unguided play. Additionally, the study highlights the importance of integrating GenAI in educational contexts to empower parents as facilitators of learning. The insights gained from the design and evaluation of BrickSmart provide valuable guidelines for developing future GenAI tools that enhance educational outcomes by supporting guided play and other interactive learning methods. Future work will explore expanding the system's capabilities and examining its application in broader educational contexts.

%% file: data/Appendix.tex
\clearpage
\appendix
\onecolumn



\twocolumn
\section{Algorithm of Brick Placement Optimization}
\label{Ap.B}

To address computational challenges, we implemented a three-stage heuristic algorithm based on the `matheuristic' approach \cite{kollsker2021optimisation}. 

\begin{figure}[h] 
\centering 
\includegraphics[width=\linewidth]{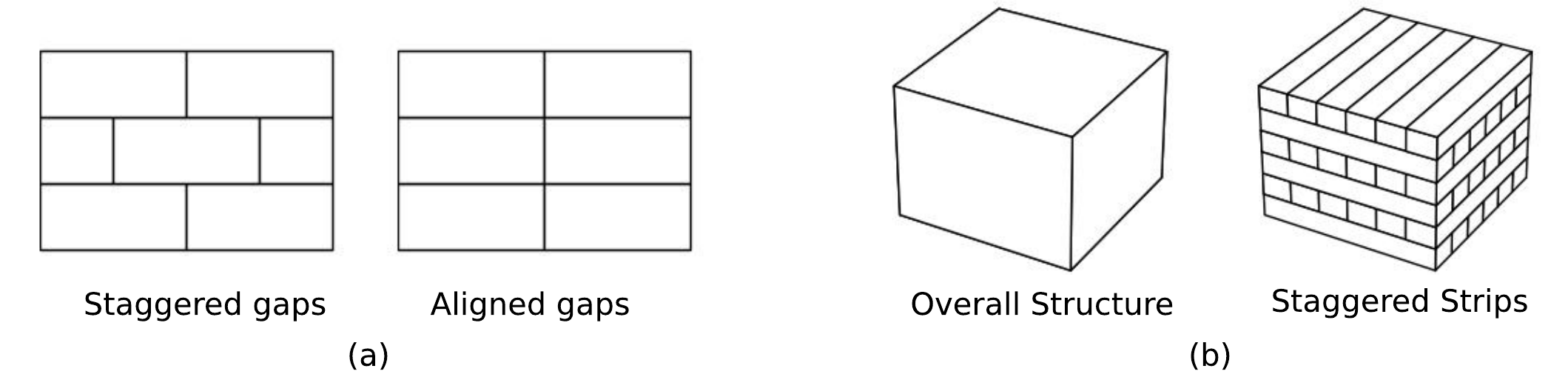} 
\caption{(a) Comparative illustration of brick placement stability: staggered versus aligned gaps. (b) Decomposition of the object into orthogonal staggered strips, layer by layer.} 
\label{fig:algorithm} 
\end{figure}

The first stage involves decomposing the 3D object into $1 \times X$ strips processed layer by layer, as shown in Figure \ref{fig:algorithm}. This reduction from three to two dimensions significantly speeds up computations, facilitating real-time instruction generation. 

The second stage uses a 1D-heuristic algorithm to segment co-planar strips into the minimum number of bricks, incorporating gap information $G$ to optimize placement:
\begin{equation}
F_{cost}(b, L) = M(l_b, l_0) + N(L, l_0) + D(b, G) + e
\label{eq4}
\end{equation}
Here, $b$ represents the brick selected in the current step, with $l_b$ denoting the brick's length. The length $l_0$ corresponds to the longest brick available, and $L$ represents the remaining length of the target strip. The multiplicity function: 

\begin{equation}
M(l_b, l_0) = \frac{l_0}{l_b}
\label{eq5}
\end{equation}
aims to minimize the frequent use of smaller bricks. The remainder counting function: \begin{equation}
N(L, l_0) = \left\{\begin{matrix}
\left \lceil  \frac{L-\rho}{l_0} \right \rceil + LpInt(\rho), \quad L \ge \rho
 \\
LpInt(L), \quad L < \rho
\end{matrix}\right.
\label{eq6}
\end{equation}
where $\rho$ determines when to engage integer linear programming for the remaining part. The optimization is described as follows, similar to the binary programming model: 
\begin{equation}
min\ N = \sum_{b^*\in \mathcal{B^*}}^{} x_{b^*}, s.t. \sum_{b^*\in \mathcal{B}^*}^{} l_{b^*} x_{b^*} = L, \quad x_{b^*} \in \mathbb{Z}^+, \forall b^*\in \mathcal{B}^*
\label{eq7}
\end{equation}
The border-gap evaluation function is defined as: \begin{equation} D(b, G) = \gamma_1 \exp(-\gamma_2 d_{b,G}) \label{eq10} \end{equation} where $d_{b,G}$ is the closest vertical distance between the borders of brick $b$ and the recorded gaps $G$, and $\gamma_1$, $\gamma_2$ are exponential coefficients. The term $e$ represents a random perturbation, introducing variability into the algorithm to avoid local minima and enhance solution diversity. This holistic approach addresses both efficiency in brick usage and aesthetic considerations by considering gap placements within the overall structure.

\vspace{10pt}

In the final stage, adjacent bricks are merged to form larger units, reducing the number of components and simplifying the building instructions. For example, two adjacent \(1 \times 4\) bricks are combined into a single \(2 \times 4\) brick. 

Algorithm \ref{alg1} provides a detailed description of the entire three-stage procedure. This ensures that the building process is both efficient and user-friendly, addressing the challenges of computational complexity while adhering to practical construction techniques.

\begin{algorithm}[h]
\caption{Brick Placement Optimization}
\begin{algorithmic} 
\label{alg1}
\STATE {\bf Input:} A matrix of overall voxelized structure \textit{voxel\_matrix} and a list of brick size \textit{brick\_list}
\STATE{\bf Output:} A list of bricks to make up the structure
\STATE \textit{strip\_list, gap\_list} $\gets$ SEGMENT(\textit{voxel\_matrix})
\STATE \textit{build\_list} $\gets$ NULL
\FOR{\textit{strip} in \textit{strip\_list}}
    \STATE \textit{L} $\gets$ length of \textit{strip}
    \FOR{\textit{brick} in \textit{brick\_list}}
        \STATE \textit{cost} $\gets$ $\text{F}_{cost}$(\textit{brick}, \textit{L}) \\
        find the lowest \textit{cost} and corresponding \textit{brick}\\
        update \textit{build\_list}
    \ENDFOR\\
    \quad update \textit{gap\_list}
\ENDFOR
\\
\STATE \textit{build\_list} $\gets$ MERGE(\textit{build\_list})
\\
\STATE return \textit{build\_list}
\end{algorithmic}
\end{algorithm}

\clearpage

\onecolumn

\section{Prompts for Three Steps}
\label{Ap.C}
\input{table/tab_prompt}

\clearpage

\section{Questionnaire of Spatial Language Testing}
\label{Ap.D}
\begin{figure}[!h]
    \centering
    \includegraphics[width=0.84\linewidth]{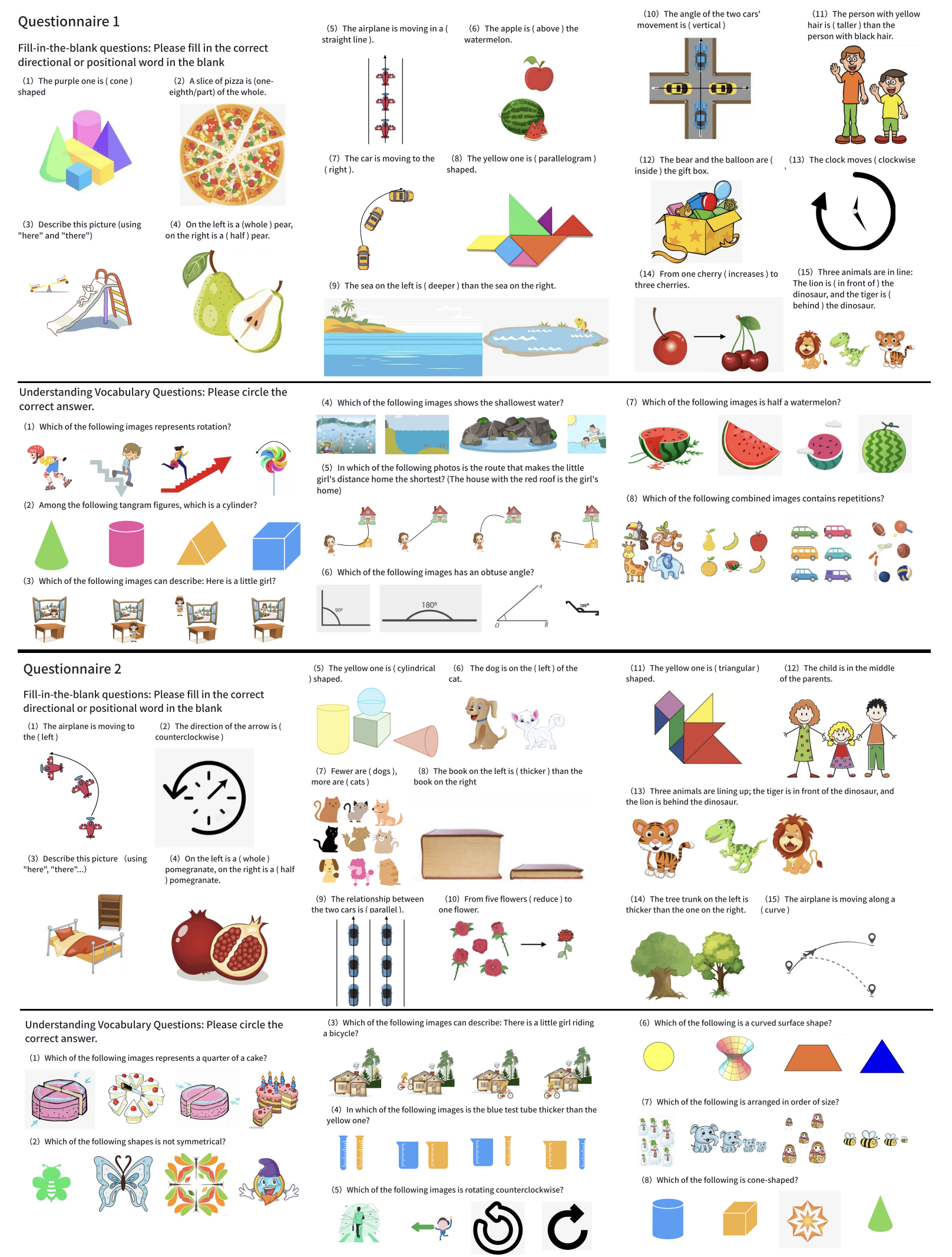}
    \caption{Questionnaire of spatial language testing. The pre-test and post-test questionnaires are alternated between Questionnaire 1 and Questionnaire 2.}
    \label{fig:question}
\end{figure}

%% file: table/tab_prompt.tex
\begin{table*}[h!]
\centering
\small
\caption{The detailed prompts BrickSmart employs across three steps. Each step is tailored to enhance children’s spatial reasoning and language skills through structured interactions and tasks.}
\begin{tabular}{l|p{0.9cm}|p{12.5cm}}
\toprule
\textbf{Steps} & \textbf{Goal} & \textbf{System Prompts}\\ \hline
\multirow{2}{*}{Step 1} & Generate descriptions & Your task is to break down the story and scenes described by the child into several describable 3D objects, and output them sequentially into a structured string list object\_list. For example, if the child says "A monkey with big eyes is climbing a tree," the output string list should be: object\_list= Monkey, big eyes, action is climbing a tree, Tree. Please only break down and output based on the child’s description without adding extra information or your own ideas. If the description lacks sufficient detail, output an empty string list object\_list. You already have an initial list: \{object\_list\}. Based on new descriptions or existing content, you need to refine and complete the entries in the list. Here is the conversation history so far: \{chat\_history\}. \\ \cline{2-3}
& Generate prompts for 3D models& 
You are an expert at crafting text prompts for generating 3D models, specializing in transforming children's imaginative words and narrated diaries into delightful, cartoon-style 3D models. Your prompts should focus on describing a single object rather than a scene, ensuring that the description is suitable for conversion into LEGO models. When rewriting user input, consider the following: The generated 3D model should avoid unnecessary details, should have a suitable center of gravity, and should clearly and independently represent a single entity. The appearance can be slightly enhanced to appeal to the aesthetic preferences of 6-8-year-old children. 
Ensure the model is suitable for construction using only basic LEGO bricks, emphasizing square and rectangular forms. Your output should be a detailed sentence or a list of descriptive words separated by commas, written in English.
\\ \hline
\multirow{3}{*}{Step 2} & Under-standing tutorial images & Based on the provided LEGO assembly tutorial image, thoroughly analyze and describe the current building task, outputting a string parameter instruction: (1) Describe the overall structure and design features of the LEGO model shown in the image. (2) Identify and describe the types of LEGO pieces and their colors appearing in the image. (3) Outline the assembly steps shown in the image, including any special assembly techniques or details that require special attention. (4) Use professional LEGO terminology to enhance the accuracy and professionalism of the description. (5) Ensure the text description accurately reflects the image content, with clear, professional, and detailed language to facilitate understanding of the assembly process.\\ \cline{2-3}
& Vocabu-lary selection & You are a spatial language teacher, responsible for selecting spatial language tasks based on the task and student's proficiency. Use the following information: LEGO assembly tutorial description: \{instruction\}. Eight spatial language dimensions: ...  
User's current spatial language proficiency: \{understand\_level\}, representing progress in each dimension (as a percentage). Select 3 suitable spatial language categories, focusing on those relevant to the LEGO task and where proficiency is lower. Output a list spatial\_list with 3 elements, each from 0 to 7, representing category indices.\\ \cline{2-3}
& Guidance generation & You are a family guide helping parents improve their child's spatial language. Based on the current LEGO tutorial, generate real-time prompts for parents to teach these words: Word: {word\_1}, Stage: \{stage\_1\}; Word: \{word\_2\}, Stage: \{stage\_2\}; Word: \{word\_3\}, Stage: \{stage\_3\}.
The current tutorial is: \{instruction\}, showing a top view (current step) and a whole view (completed so far). Use the building task to teach the words, matching each to its learning stage. Example format: Word: Circle, Stage: Noun Explanation. Prompt: Explain that a circle has no corners, and all points on the edge are equally distant from the center. Example: Point out circular bricks or designs to help understanding.\\ \hline
\multirow{2}{*}{Step 3}& Guidance generation & You are an assistant helping parents improve their child's spatial language skills. They’ve built LEGO models: \{objects\}. Your task is to guide parents to move these models and have the child describe the actions to enhance spatial understanding. Output format: Vocabulary: The word to learn. Movement Example: Specific movement of the object. Parent Prompt: Example guidance for parents. Example: Vocabulary: Left/Right. Movement Example: Move the figure forward, then turn left. Parent Prompt: "Look, the figure turns left. Can you make it turn right?".  For the {num\_words} keywords: \{keywords\}, provide suggestions. \\ \bottomrule
\end{tabular}

\label{tab:Prompts}
\end{table*}